\documentclass[twocolumn,amsmath,amssymb,prl,english]{revtex4}

\makeatletter

\usepackage{graphicx}
\usepackage{amsmath}
\usepackage{amssymb}
\usepackage{babel}
\usepackage{color}
\makeatother

\usepackage{graphicx}
\usepackage{dcolumn}
\usepackage{bm}

\begin{document}

\title{Photon energy lifter}
\author{Zeno Gaburro}
%\email[]{gaburro@science.unitn.it}
\author{Mher Ghulinyan}
\author{Francesco Riboli}
\author{Lorenzo Pavesi}
\affiliation{Department of Physics, University of Trento and
CNR-INFM, Povo, I-38050 Trento, Italy }

\author{Alessio Recati$^a$, and Iacopo Carusotto}
\affiliation{CRS BEC-INFM, Povo, I-38050 Trento, Italy}
\affiliation{$^a$ also at ECT*, Villazzano, Trento, Italy}

\begin{abstract}
We propose a time-dependent photonic structure, in which the
carrier frequency of an optical pulse is shifted without changing
its shape. The efficiency of the device takes advantage of slow
group velocities of light attainable in periodic photonic
structures. The frequency shifting effect is quantitatively
studied by means of Finite Difference Time Domain simulations for
realistic systems with optical parameters of conventional silicon
technology.
\end{abstract}

\maketitle

In several applications, particularly in the field of optical
signal processing, it is desirable to dispose of techniques for
continuous and efficient photon energy conversion. Most of the
nowadays available techniques rely on inelastic optical processes.
For example, stimulated Raman scattering has been successfully
employed for frequency conversion and lasing\cite{siliconlaser}.
Unfortunately inelastic processes are usually not efficient, often
requiring high intensities. Finding new approaches to such a task
is a challenging problem. Recently Reed {\sl et
al.}\cite{shockwave} have proposed a way to change the color of
light by capturing it on a propagating shock wave front and
successive reemission.

Dynamical photonic structures, in which the optical constants of
the medium can be modulated while the light is propagating inside
it, have been discussed very recently (see, e.g.,
\cite{stoppingFan} and references therein). Structures of this
kind, originally suggested for stopping the light, open up an
intriguing research area, in particular, for storing  or
manipulating optical information. In this Letter we face the
photon energy conversion problem exploiting the simple idea of
changing the photon frequency using materials with time-dependent
optical properties. We also suggest a realistic implementation of
this idea, by means of a photonic structure which can be realized
using nowadays available technologies. The proposed device
provides an intriguing possibility to shift continuously the
individual energies of photons. We wish to emphasize this aspect
by referring to it as {\sl photon energy lifter}.

It is a simple consequence of translational invariance that the
wavevector $k$ of an electromagnetic wave propagating in a
homogeneous and non-absorbing dielectric with a time-dependent
refractive index $n(t)$ is a conserved quantity. As the frequency
is related to the wavevector and the refractive index by $\omega=
\frac{c}{n}k$, where $c$ is the vacuum light speed, a change of
refractive index by $\delta n$ results in a change of the
frequency of a plane wave by:
\begin{equation}
\frac{\delta\omega}{\omega}=-\frac{\delta n}{n+\delta n}.
\end{equation}
The energy required for the frequency change is provided by the
work done while changing the refractive index of the medium.

This effect can be used in a photonic device to shift the carrier
frequency of a wave-packet (light pulse), without dramatically
affecting the pulse shape: to do this, it is enough to trigger the
refractive index shift once the pulse has entered the medium, and
to complete it before the pulse starts exiting. In this way, the
travelling pulse is fully contained in the medium during the whole
index shift process, and the medium effectively acts as an
infinite one. The optical linearity of the medium guarantees that
the different $k$-components are decoupled, and the pulse shape is
preserved in both the real and the $k$-space.

Unfortunately, the experimental implementation of this set-up is
very demanding for available technology. As the travel time
$\tau_{travel}=L/v_g$ of the pulse across the structure must be
longer than the pulse duration $\tau_{pulse}$ plus the switching
 time $\tau_{sw}$, the device length $L$ has to be longer
than $L_{min}=v_g(\tau_{pulse}+\tau_{sw})$. For realistic values
of $\tau_{pulse}$ and $\tau_{sw}$ of the order of a few tens of
picoseconds, and group velocities $v_g$ of the order of a
significant fraction of $c$, $L_{min}$ can be as long as some
millimeters. Such a length it is not only undesirable in the
perspective of miniaturization and integration, but, more
importantly, introduces severe limitations on the performances of
the system. If the index change is achieved by electro-optic
effects, long electrodes would imply too slow RC time constants;
if, on the other hand, the index change is achieved by optical
effects (such as photocromism), the required optical power would
become very high for practical applications.

It becomes apparent that one should increase the travel time
across the structure. We are therefore led to investigate periodic
photonic structures with particular dispersion relation, in which
the group velocity is reduced significantly \cite{notecav}.

Various photonic systems have been recently investigated to slow
down the group velocity of a travelling optical pulse. Among them
line-defect photonic crystals \cite{Krauss} and coupled resonator
optical waveguides (CROW) \cite{Yariv, APL_slow} are the most
promising. In this Letter we focus ourselves on the investigation
of a CROW structure to demonstrate the concept of a photon energy
lifter.

\begin{figure}[t!]
\centering
\includegraphics[width=8.5cm]{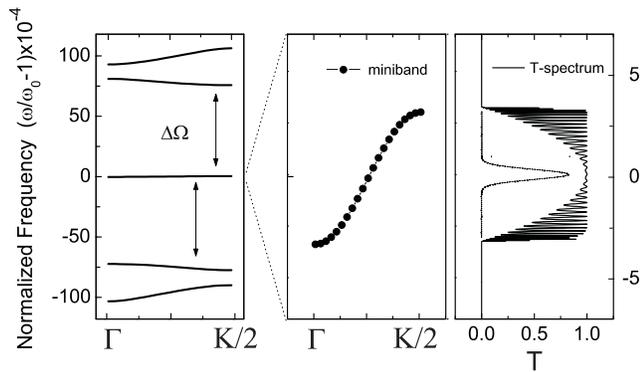}\\
\caption{Band structure of the 1D model of the CROW (left panel).
The arrows indicate the frequency separation $\Delta\Omega$
between the miniband and the nearest allowed photonic states.
Enlarged view of miniband (central panel). Trasmission spectrum
and spectral shape of the signal pulse, chosen to fit in the
almost flat region of the spectrum (right panel). For the actual
parameters see text.} \label{miniband}
\end{figure}

A CROW is a spatially periodic many-cavity system, in which the
degeneracy of the resonant states is lifted by the coupling
between neighboring cavities. As a result, a miniband is created
(see Figure \ref{miniband}), with a dispersion law $\omega(k)$
whose width is proportional to the coupling
strength~\cite{coupledcav} and whose group velocity is orders of
magnitude lower than the one associated to the average refractive
index ${\bar n}$ of the structure. This is a consequence of the
strong light localization in the cavity states. For example the
group velocity at the center of the miniband shown in Figure
\ref{miniband} is two order of magnitude smaller than the vacuum
light speed.

In a infinite CROW system the discrete translational symmetry of
period $\ell$ guarantees that the quasi-wavevector $k$ (defined
modulo integer multiples of the reciprocal lattice vector
$\kappa=2\pi/\ell$) is conserved under any refractive index
modulation which preserves the discrete translational symmetry. A
time-dependence in the refractive index will give a time
dependence of the dispersion law. Provided the electromagnetic
field dynamics is limited to the miniband states only, the
frequency of the wave is fixed by the miniband dispersion law.
Thus the change in the refractive index results in a corresponding
change of the carrier frequency of the wavepacket.

There is a basic limitation to the maximum speed at which the
refractive index can be changed. Indeed, to preserve the coherent
information, the pulse should remain in the miniband during the
whole process. This is the case provided that the refractive index
tuning is adiabatic \cite{Galindo}. As an estimate, the tuning
time $\tau_{sw}$ should be much larger than the inverse of the
minimum frequency separation $1/\Delta\Omega$ between the miniband
and the nearest allowed photonic states, shown schematically in
Figure~\ref{miniband}.

In a finite CROW system, for which the impedances are carefully
matched at the input and output interfaces, the central part of
the miniband corresponds to a frequency window in the
transmittivity, which is flat and almost unity, Figure
\ref{miniband}. The travel time $\tau_{travel}$ is fixed by the
group velocity \cite{flatvg}. The transmittivity profile
guarantees that an incident pulse, whose energy dispersion resides
in the flat part of the spectrum, penetrates and crosses the
photonic structure without significant distortion. The strong
reduction of $v_g$ with respect to $c$ implies a much looser bound
$L_{min}$ on the system length in order to be able to perform the
refractive index modulation while the pulse is contained in the
structure.

\begin{figure} [t!]
\centering
\includegraphics[width=7.5cm]{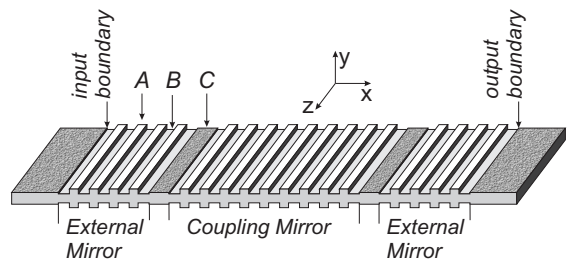}\\
\caption{ On scale, for clarity, simplified schematic representation of the
CROW structure. It consists in representing external dielectric mirrors
with only four periods instead of 13, nine periods instead of 27
in the coupling mirrors, and on reporting two cavities instead of
45.
% A coupling-to-environment strategy is chosen by the use of
%weaker external mirrors, providing a balanced compromise between
%localization and light transmission and resulting in a broad and
%high-transmission miniband spectrum. The structure is
%SiO$_2$/Si/SiO$_2$ (symmetric) in the y direction and extends to
%infinity in the z direction.
Light travels in the positive x
direction. }\label{structure}
\end{figure}

In order to illustrate the idea, we perform Finite Difference Time
Domain (FDTD) simulations of a pure one-dimensional (1D) structure
composed by 45 Fabry-Per{\'{o}}t microcavities, coupled to each
other through 27-period  Distributed Bragg Reflectors (DBRs). The
mirrors and the cavities are constituted by quarter- and
half-wavelength layers, respectively, and are centered at
$\lambda_0=1550$~nm. Moreover the input-output mirrors are
suitably designed in order to have a good impedance matching
between the finite-1D photonic structure and input/output channels
(see Figure~\ref{structure}). Assigning letters $A$ and $B$ to the
mirror layers and $C$ to each cavity layer, we choose their
refractive indices equal to $n_A=3$, $n_B=2.426$ and $n_C=2.688$,
respectively. The choice of these values is due to the realistic
implementation with the available silicon technology we suggest in
the second part of this Letter.

\begin{figure}
\centering
\includegraphics[width=7.5cm]{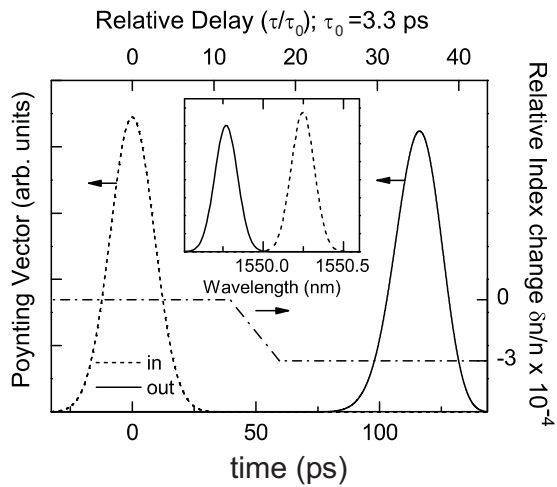}\\
\caption{FDTD simulation of the envelope of Poynting vectors at
the input (dashed line, positive entering) and output boundary
(solid line, positive exiting) of the coupled-cavities structure.
Input and output boundaries are defined in Figure~\ref{structure}.
The refractive index of all the layers is time-dependent, as shown
by the right y-axis. Upper time scale is relative to the delay
($\tau_0$) in a homogeneous medium with the average refractive
index of the 1D-structure. The inset shows the spectra of the
input and output pulse electric fields. }\label{ccavtime}
\end{figure}

The time dependence of the structure in the simulation consists in
a relative shift $\delta n/n$ of the refractive index, applied to
the whole structure, triggered after the injection of the pulse
and completed before the pulse exits.  For simplicity, a linear
time dependence of $\delta n/n$ has been considered. The number of
microcavities has been determined assuming that the index change
can be completed within 20 ps. Consistently, the minimum tuning
time $\tau_{sw}$ for being adiabatic is, for our parameters, much
smaller, of the order of 10 fs. The final device length is about
370 $\mu$m. A switching time longer than 20~ps would simply imply
a proportionally longer structure. A value $\delta
n/n=-3\times10^{-4}$ has been chosen on the basis of realistically
attainable values in semiconductor technology
\cite{electropticmaterials}. It is worth noticing that
significantly larger index shifts, hence, frequency tuning, would
be  possible, just by resorting to electro-optic materials such as
LiNbO$_3$ \cite{electropticmaterials} .

The FDTD simulation of the pulse transport is reported in
Figure~\ref{ccavtime}. The output signal (solid line) is plotted
versus both time and its relative delay to the travelling time
($\tau_0$) of a pulse travelling in a structure with the same
length, but consisting of a single homogeneous layer with the
average refractive index of our structure. The pulse delay is
116~ps, which is larger by a factor of $\sim35$ compared to the
homogeneous case. The resulting vacuum wavelength shift $2\pi
c/\delta\omega=-\delta\lambda\simeq -0.49$~nm is shown in the
inset, and corresponds to the expected value
($\delta\lambda/\lambda_0\simeq\delta n/n$). Finally, as expected,
the distortion of the input pulse is negligible and the efficiency
of conversion is larger than 95\%.
\begin{figure}[t!]
\centering
\includegraphics[width=7.5cm]{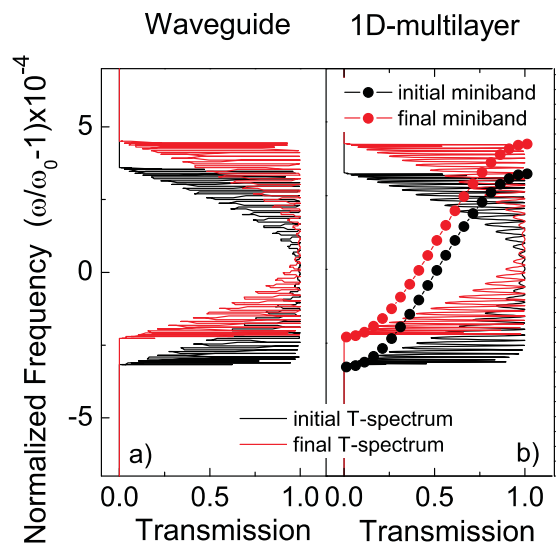}\\
\caption{ (color online) A comparison between the calculated
transmission spectra of the initial and shifted minibands in the
case of a waveguide (a) and a 1D-multilayer structure (b). The
miniband dispersion (full circles) in the 1D case is also plotted
in (b).} \label{spectrum}
\end{figure}

A realistic implementation of the photon energy lifter with
available standard technology is based on silicon waveguides.
Moreover, as it will become clear later, such a system can be
suitably described by the 1D system previously used and analyzed
to illustrate the photonic energy lifter principle. The typical
architecture of a silicon waveguide working in the infrared range
is composed by a silicon guiding layer of about 260nm thickness,
sandwiched between SiO$_2$ cladding layers. The index contrast
between the two materials assures light confinement in the silicon
core. The photonic structure (DBRs and cavities) can be realized
modulating the thickness of the silicon slab, which corresponds to
an effective refractive index modulation of the fundamental
transverse mode, Figure~\ref{structure}. We perform, using an
eigenmode expansion method \cite{sudbo}, a full two dimensional
simulation for the slab waveguide before and after the index
switching. For the sake of computational simplicity, as sketched
in Figure~\ref{structure}, we assume a vertical symmetric (the
$y$-direction) architecture. The physical parameters of the
integrated device (length and thickness of the layers) -- which
eventually provide the refractive indices used in the 1D FDTD --
have been chosen in order to maintain a good modal matching
between input/output waveguides, cavities and DBRs. In this way
the insertion losses in the resulting two dimensional CROW slab
device appear to be negligible, of the order of $0.1$dB. In Figure
\ref{spectrum} both initial and shifted miniband spectra of the
slab structure (panel (a)) and the 1D model of the CROW (panel
(b)) are shown. The agreement is quite satisfactory, confirming
that the 1D structure (and the 1D FDTD code) is a good description
for the waveguide system.

Up to now the basic idea of the photon energy lifter was based on
the hypothesis that the refractive index can be simultaneously
changed in the whole structure while the pulse is travelling
across it. Another possible scheme, which can simplify the
experimental realization is what can be called a travelling-wave
design \cite{tavellingwave}. In this scheme, the electrical field
for the electro-optic effect is not applied simultaneously to the
whole device, but is injected as a microwave from one end of the
electrode pair. Electrodes work as a transmission line. In this
design, the electrode capacitance, being distributed, does not
limit the modulator speed, and the electrodes can be made very
long.

Since the photonic structure under study is designed for
\emph{optical} wavelengths, it behaves as a homogeneous medium at
microwaves.
%Since the size of the photonic features is of the order of the
%\emph{optical} wavelength, the photonic structure behaves as an
%essentially homogeneous medium at microwaves.
Therefore, it is expected that the group velocity at microwaves is
determined by the average refractive index, $n_{\mu w}$, of the
structure and thus it is much higher than the group velocity of
the optical pulse. Moreover, the considered pulse wavelength,
being much shorter than that of the microwave, sees an almost
translational invariant system with a time-dependent refractive
index.

\begin{figure}
\centering
\includegraphics[width =7cm]{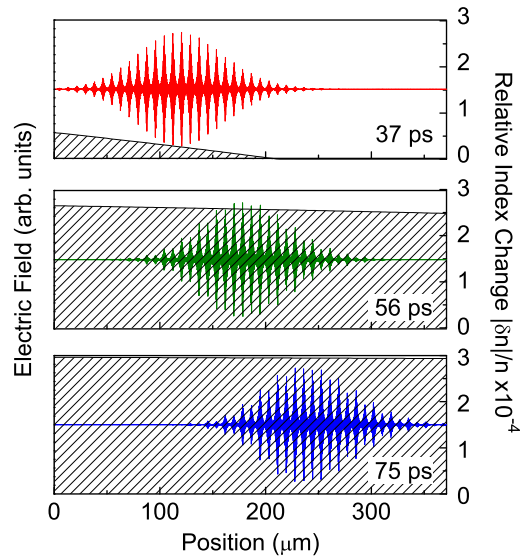}\\
\caption{(color online) Successive snapshots of the electric field
of the optical wave and of the index-driving microwave at 37, 56
and 75~ps. The differently colored electric fields have an
illustrative scope and refer to a blueshift of the input pulse
wavelength. The structure of the pulse electric field reflects the
coupled cavity structure (maxima are located in correspondence of
$C$ layers. }\label{sample}
\end{figure}

We assume, for our FDTD simulations, a conservative value of
$n_{\mu w}$=3 as the effective refractive index of the mixed
Si/SiO$_2$ structure at microwaves. Therefore the optical $v_g$ in
our structure is lower than the microwave group velocity $c/n_{\mu
w}$ by over an order of magnitude. The microwave driver signal is
taken as a smooth function $\propto(1-\exp(-t/\tau))$ with
$\tau^{-1}=100$GHz. The microwave is sent into the structure with
a 35~ps delay from the input light signal, i.e., when the
injection of the light signal is essentially completed
Figure~\ref{ccavtime}. Sample electric field snapshots resulting
from the combined optical and microwave simulations are reported
in Figure~\ref{sample}, where an instantaneous correspondence of
the microwave amplitude with the relative refractive index change
has been assumed. The results of the travelling-wave simulation
are not distinguishable from the results of Figure~\ref{ccavtime},
which validates this method.
A more practical travelling wave design would consider a device
architecture where the light is also laterally confined (along the
$z$ direction in Figure \ref{structure}), e.g., a so-called ridge
waveguide.

In conclusion, we have proposed a novel photonic device which is
able to shift the carrier frequency of an optical pulse without
affecting its shape. This is obtained by changing in time the
refractive index of a coupled-cavity system while the pulse is
propagating through the structure and is fully contained in it:
the low value of the group velocity in coupled-cavity systems is a
key ingredient for this procedure to be possible in a miniaturized
device. Full numerical simulations have been used to confirm our
predictions for realistic silicon-based systems with different
switching mechanisms.

In addition to the strong technological
interest of the present proposal for a photon energy lifter, it
is another example of the intriguing properties that electromagnetic waves
have when travelling in a time-dependent photonic structure.

We acknowledge the financial support by MIUR through FIRB
(RBNE01P4JF and RBNE012N3X) and COFIN (2004023725) projects and by
PAT through PROFILL project.


\begin{thebibliography}{99}
\bibitem{notecav}
Single optical microcavity can localize strongly the light
\cite{cavities} and, in principle, are suitable for changing the
photon energy. However, these structures strongly distort pulses.
\end{thebibliography}
\end{document}